# Propagation of transition front in bi-stable nondegenerate chains: model dependence and universality

I.B. Shiroky and O.V. Gendelman*

Faculty of Mechanical Engineering, Technion

* - contacting author, ovgend@tx.technion.ac.il

We consider a propagation of transition fronts in one-dimensional chains with bi-stable nondegenerate on-site potential. If one adopts linear coupling in the chain and piecewise linear on-site force, then it is possible to develop well-known exact solutions for the front and accompanying oscillatory tail. We demonstrate that these solutions are essentially non-robust. Various approximations for the on-site potential with the same basic parameters (height and coordinate of the potential barrier, energy effect and distance between the potential wells) lead to substantially different front velocities. Besides, inclusion of even weak nearest neighbor nonlinearity drastically modifies the front structure and parameters. The energy concentration in the front zone leads to a dominance of the nonlinear term. It turns out that the dynamics can be efficiently studied in terms of an equivalent model with a single degree of freedom. This estimation leads to accurate prediction of the front velocity and parameters of the oscillatory tail. Moreover, it turns out that the solution is robust - exact shape of the on-site potential weakly effects the front parameters. This finding also conforms to the simplified model, since the latter invokes only the general shape characteristics of the on-site potential.

1. Introduction

In condensed matter, many processes accompanied by energy release, occur through propagation of transition fronts in the bulk at a nanosecond scale [1, 2, 3, 4]. Estimation of the characteristics of such transitions, especially of the front velocity and structure of the oscillatory tail, are of considerable interest in theoretical and experimental studies.

It is beneficial in many cases to invoke discrete models for description of these phenomena. In such models, the transition front (or other structural defect) is to overcome the potential barrier caused by the discrete lattice. For this sake, it requires either external forcing or energy release in the lattice, related to the front propagation. This micro- scale effect disappears in continuous models, where the solutions, which represent defects, can move freely without drag. The discrete models, on the other hand, in principle allow derivation of a relation between the external force or energy gain and the front velocity, commonly known as the "kinetic relation". At the micro level, when no dissipation is assumed, the effect, which describes the removal of the released energy, is known in literature as "radiative damping" [5]. The lattice waves that accompany the transition front have to remove at least the major portion of the gained energy, while the rest of the energy may be spent for creating new surfaces as it happens in the case of crack initiation [6, 7].

Analysis of the discrete models with multi-equilibria potentials of interaction goes back to early work of Dehlinger [8] and dynamical model of Frenkel and Kontorova [9]. Various derivatives of these models appeared throughout the years to describe different phenomena. Classic examples of this sort are dislocations in metals [3, 4], lattice distortions around twin boundaries [10], domain walls in ferro-electrics [1], and crack propagation [11]. Truskinovsky and Vainchtein in [2] address martensitic phase transitions by presenting a discrete model with long- range interactions. The model allows the derivation of a macroscopic dissipation law specified as a relation between force and velocity. The dissipation is due to radiation of lattice waves that carry energy from the front. Dynamics of crowdions in anisotropic crystals was studied in [12, 13]. In [14] a damped and externally driven FK chain is studied and the threshold forcing amplitude is found. This model may be beneficial in describing a chain of ions trapped in a metallic surface with an external AC electric field as the drive, as well as in the study of Josephson junctions. Other applications can be found in the comprehensive review of Braun and Kivshar [4].

A common configuration that has been adopted in many works is that of a bi-stable on-site potential [15, 16, 5]. The most interesting case is the non-degenerate on-site potential with certain energetic difference between the minima; this difference dictates the direction of the front propagation. The linearly coupled chain with a bi-stable on-site potential was studied in previous works. The simplest model where both wells have same curvature was proposed in the seminal work of Atkinson and Cabrera [3] and further explored in the early works of Ishioka [17] and Celli [18] . The advantage of this potential is that an analytical solution can be obtained through direct Fourier transform. The technique was further implemented in [2, 5, 19]. The case with different well curvatures requires application of the Wiener-Hopf method. It was done for systems with asymmetric parabolic double well in [20, 21, 22]. Similar models with nonlinear coupling received less attention. Recent studies [23, 24] address the case of generic coupling with on-site dissipation and suggest a law that connects the transported energy with the velocity and dissipation ratio.

The studies mentioned above widely explored the models with linear and piecewise linear interactions. The reason is obvious – a possibility of exact solutions and comprehensive exploration of the transition front and the oscillatory tail. However, even if one restricts himself by the one-dimensional setting and considers a model with minimal possible set of interactions (for instance, only the on-site and nearest – neighbor), still a robustness of the linear approximation is rather questionable. In this work, we are going to demonstrate that the linear approximation is indeed not robust in a sense that the addition of small nonlinearity leads to qualitative changes in the structure of the propagating front and the oscillatory tail. Besides, even if the realistic system is reduced to the one-dimensional model, the exact shapes of the interaction potentials are hardly known and one can operate only with some particular characteristics. Commonly, only such basic parameters as the energy barrier, energy effect of the transition, the distance between the minima and, sometimes, the frequencies of oscillations in the potential wells can be measured or simulated. Any specific set of these parameters can correspond to uncountable infinity of possible approximations of the potential functions. We are going to explore how the specific choice of such approximation affects the transition front characteristics. It turns out that the effect is noticeable in the case of the linear nearest-neighbor interaction and almost absent if the nonlinearity is taken into account.

2. **Description of the model and the case of the linear nearest-neighbor interaction**.

The basic model is adopted from [15, 5] where the problem of a chain with the linear inter-particle interaction and an on-site bi-parabolic potential is addressed. Here we expand this model in two ways. First, we include a cubic nonlinear inter-particle coupling. Second, we don't specify the exact shape of the on-site potential, but only fix its main characteristics. The bi-parabolic shape will be a particular case that we will refer to later. The Hamiltonian of the system is given by Eq. (1).

$$H = \sum_{n=1}^{\infty}\left[ \frac{p_n^2}{2} + \frac{c}{2}(\varphi_{n+1} - \varphi_n)^2 + \frac{\beta}{4}(\varphi_{n+1} - \varphi_n)^4 + U(\varphi_n) \right] \qquad (1)$$

$\varphi_n$ is the displacement of the $n^{th}$ particle from the initial equilibrium state (meta-stable). $c$ and $\beta$ are respectively the stiffnesses of the linear and nonlinear nearest-neighbor springs, $p_n = \dot{\varphi}_n$, the mass of each particle is set to unity. The shape of the nondegenerate bi-stable on-site potential $U(\varphi)$ is defined by the energetic effect $Q$, the height of the potential barrier $B$, coordinate of the barrier $b$ and coordinate of the stable state $\varphi^*$. Three examples that conform to these requirements are shown in Figure 1. Position of the meta-stable state is set to zero. In this setting, the energetic effect $Q$ is the driving force which determines the favorable direction of the reaction.

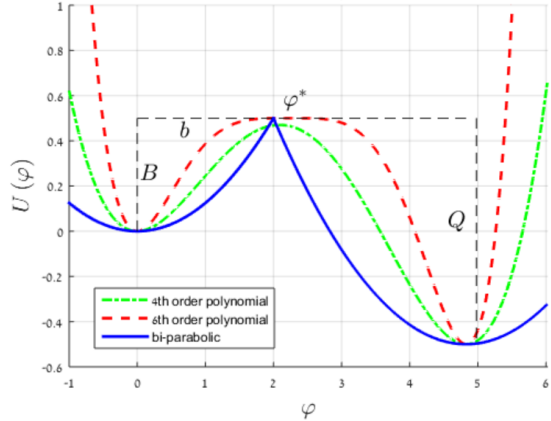

*Figure 1 - On-site nondegenerate potential $U(\varphi)$. Tree possible approximations with the same general shape parameters are presented: solid-blue-bi-parabolic potential, line-dotted green – 4$^{th}$ order polynomial, dashed red – 6$^{th}$ order polynomial.*

Without loss of generality the linear coupling coefficient is set to unity; $c=1$. The special case of system with Hamiltonian (1) with linear inter-particle interaction ($\beta=0$) and a symmetric bi-parabolic potential with equal curvatures $\omega_0$ is studied in [5]. Linear dispersion relation for this model is given by:

$$\omega^2 = 4\sin^2\left(\frac{k}{2}\right) + \omega_0^2 \qquad (2)$$

Here $\omega_0$ is a frequency of particle oscillations in each of the two potential wells. The typical dynamic response in this case is shown in Figure 2 for certain time instance. The only nonzero initial condition is the velocity of particle #1 - $\dot{\varphi}_1(0)=10$. From here on, this condition is denoted in figures as "impulse 10". It is seen that about 5 particles are within the transition area simultaneously.

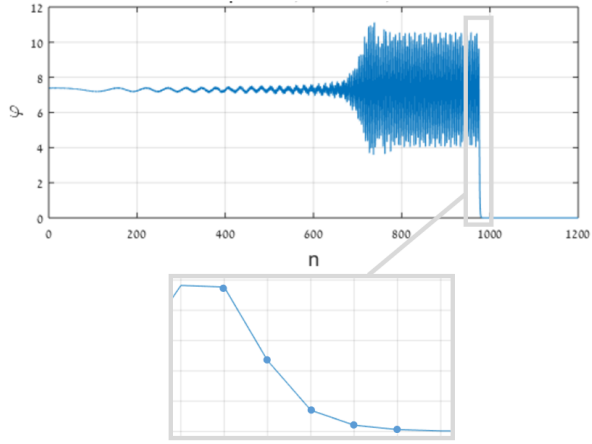

Figure 2 – Particle displacements in the chain with linear coupling. at $t=1000$. Detailed structure of the front zone is presented in the inset. Parameters: $Q/B=6, \omega_0=0.5$, initial conditions: impulse 10

For stationary propagation of the transition it is necessary that the front velocity will be equal to the phase velocity of the accompanying oscillatory tail, $V_f = V_{ph}$. It is shown in [5] that the phase velocity can be expressed analytically as a function of the driving force $Q$ through the following relationship:

$$\sum_{k \in M^+} \frac{1}{k \sin k - 4\sin^2\frac{k}{2} - \omega_0^2} = -\frac{2}{\omega_0^2} \frac{1}{1+\sqrt{1+\frac{Q}{B}}} \qquad (3)$$

Here, $M^+$ are the real and complex roots of dispersion relation (2) of the chain that satisfy $M^+ = \{k: L=0, \text{Im}\, k > 0\} \cup \{k: L=0, \text{Im}\, k = 0, kL_k > 0\}$ The phase velocity is implicitly determined for any set of the system parameters, as Eq. (3) can be satisfied only by a unique set of roots $M^+$ that are found from the dispersion relation for a single value of frequency $\omega = \omega^*$. This frequency corresponds to a single value of real wavevector $k = k^*$ Then, one obtains:

$$V_f = V_{ph} = \omega^* / k^* \tag{4}$$

To examine robustness of solution (3)-(4) to variations in the shape of the on-site potential, we numerically integrate the evolution equations obtained from Hamiltonian (1), with $\beta = 0$ different $U(\varphi)$ that have common general shape ($B, Q, \varphi^*, b$), as explained in Figure 1. The front velocity as a function of $\varphi^*$ is presented in Figure 3. For three different potentials: bi-parabolic, 4th order polynomial and 6th order polynomial, we obtain different results for the front velocity. For the 6th order polynomial, the front propagation is not observed in some range of parameters, for which it is observed for other approximations for the potentials. This finding itself indicates a significant non-robustness of the linear chain to variations in on-site potential shape.

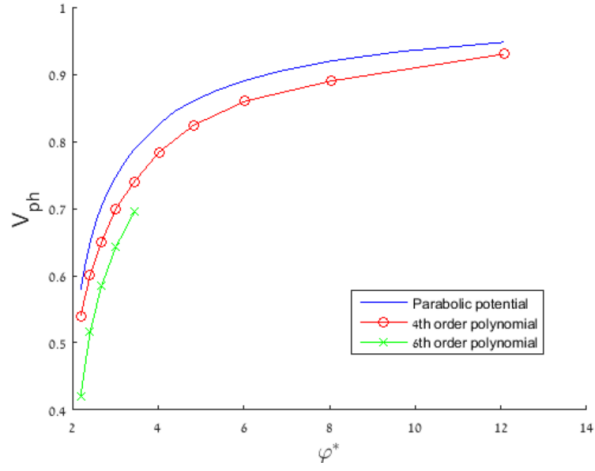

*Figure 3 – Phase velocity as a function of $\varphi^*$; Solid blue – bi-parabolic potential, 'o' red – 4th order polynomial potential, 'x' green – 6th order polynomial potential; parameters: $Q/B = 1$*

### 3. The nonlinear nearest-neighbor coupling.

Let us now consider the chain with Hamiltonian (1) and nonlinear nearest-neighbor stiffness $\beta > 0$. A typical dynamical response is shown in the $n - \varphi$ plane for fixed time instance (Figure 4), and in the $t - \varphi$ plane for one of the chain particles (Figure 5). All parameters and initial conditions besides β are similar to those used for the simulation presented in Figure 2.

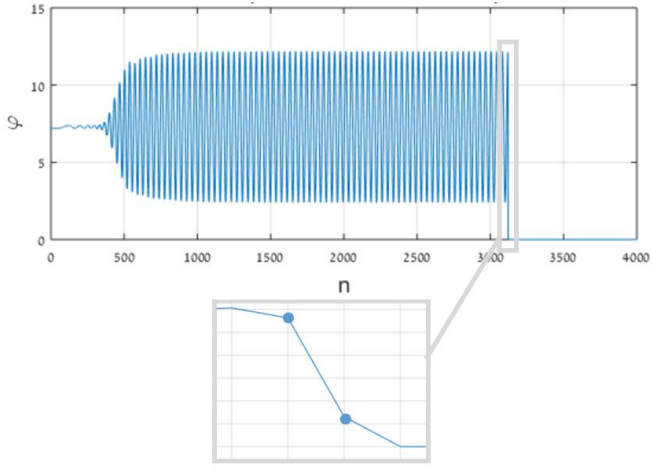

*Figure 4 - Dynamic response of the chain with cubic coupling; $\varphi = \varphi(n)$ at $t = 1000$. Parameters: $Q/B = 6, \omega_0 = 0.5, \beta = 0.2$, initial conditions: impulse 10*

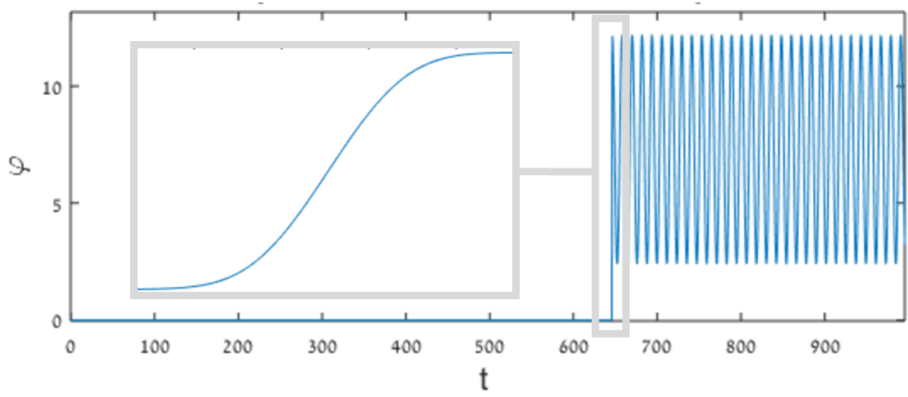

*Figure 5 - Dynamic response of the nonlinear chain; $\varphi_{2000} = \varphi_{2000}(t)$; Parameters: $Q/B = 6, \omega_0 = 0.5, \beta = 0.2$, initial conditions: impulse 10*

One immediately observes that the structure of the solution strongly differs from the case of linear interparticle interaction, even for a small nonlinearity. The transition front is considerably accelerated. The gradient at the transition area is extremely steep, with at most 3 particles in the front zone (see zoom in Figure 4). Besides, the oscillatory tail has very large wavelength. To gain further insight in the front structure, we numerically evaluate the average contributions of the quadratic and quartic terms of strain energy in the chain with the help of the following equations:

$$\bar{e}_l(n) = \frac{1}{\tau}\int_{t_1}^{t_1+\tau}\frac{(\varphi_n(t)-\varphi_{n-1}(t))^2}{2}dt; \quad \bar{e}_{nl}(n) = \frac{\beta}{\tau}\int_{t_1}^{t_1+\tau}\frac{(\varphi_n(t)-\varphi_{n-1}(t))^4}{4}dt \tag{5}$$

Here $\tau = 1/V_{ph}$ is the characteristic time of transition, $\bar{e}_l$ and $\bar{e}_{nl}$ are the average distributions of linear and nonlinear portions of interaction energy respectively. The numeric results are presented in Figure 6. For both quadratic and quartic components of the coupling energy, the concentration is extremely high in the narrow transition area compared to the rest of the chain. In addition, one notes that the quartic term is responsible for about 6 times more energy concentrated in the front than the quadratic term. The characteristic time that is needed for the solution to acquire its steady state front velocity is rather large compared to the case of the linear chain (Figure 7). This can be explained by high amount of energy concentrated in the transition zone; a lot of time is required to accumulate this amount of energy in the transition zone and to react the steady-state propagation. In conclusion, it is clear that the nonlinear interaction term qualitatively modifies the transition front, and the latter cannot be considered as perturbation of the exact solution available for the linearly coupled chain.

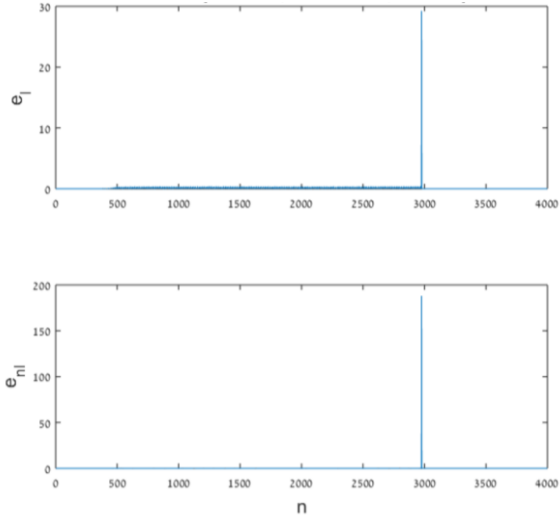

Figure 6 – Nearest-neighbor interaction energy of the chain with cubic coupling term. a) linear coupling $e_l = e_l(n)$ b) nonlinear coupling $e_{nl} = e_{nl}(n)$. Common parameters: $Q/B = 6, \omega_0 = 0.5, \beta = 0.2, t = 950$, initial conditions: impulse 10

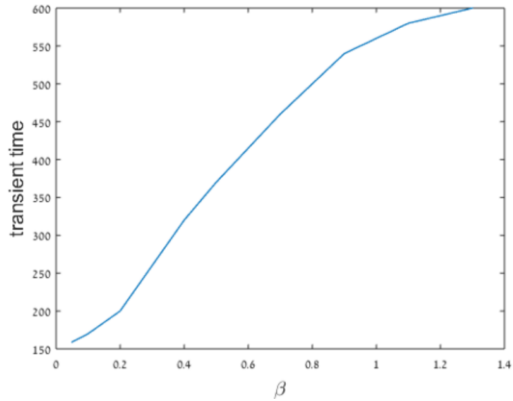

Figure 7 - Transient time as a function of $\beta$; parameters: $Q/B = 1, \omega_0 = 0.5$

### 3.1 Simplified model of the transition front.

The simplified model of the front zone for the case of nonlinear coupling is based on three observations. The first is the dominance of the nonlinear term in the transition area compared to contributions of the linear coupling and the on-site potential. Thus, in equations of motion for the particles belonging to the transition zone we can neglect in the first approximation all terms besides the nonlinear coupling. The on-site potential will affect only boundary conditions of the obtained solution. Thus, the following approximate equations of motion are considered for the particles in the transition area:

$$\ddot{\varphi}_i + \beta\left[\left(\varphi_i - \varphi_{i+1}\right)^3 + \left(\varphi_i - \varphi_{i-1}\right)^3\right] = 0 \tag{6}$$

Then, the transition front is extremely narrow, and only very few particles take part in the transition simultaneously (see Figure 4). Therefore, it is necessary to consider only very few particles in Eqs. (6) Visually, one encounters 2-3 particles inside the transition zone, but sufficient description may be obtained by considering the rapid jump of a single particle from the meta-stable state to the vicinity of the stable state.

Moreover, the gradient in the transitional region is extremely steep when compared to adjacent layout within the two wells. Thus, one can adopt that the single particle inside the front is attached to a fixed particle that still has not left the metastable position $\varphi = 0$. For the steady state, the front velocity must be equal to the phase velocity of the oscillatory tail. As it follows from dispersion relation (2), very large phase velocities $V_{ph} = \omega / k$ are possible only if the wavenumber $k$ is small, i.e. close to the left bandgap (see Figure 8). The energy in the oscillatory tail is very small compared to those concentrated in the front; therefore, one can admit that the transiting particle is attached to an immobile point with coordinate $\Delta$, defined as the first maximum of the oscillatory tail behind the transition front (see Figure 9). The resulting approximate single-DOF equation for the particle inside the transition front is presented in (7).

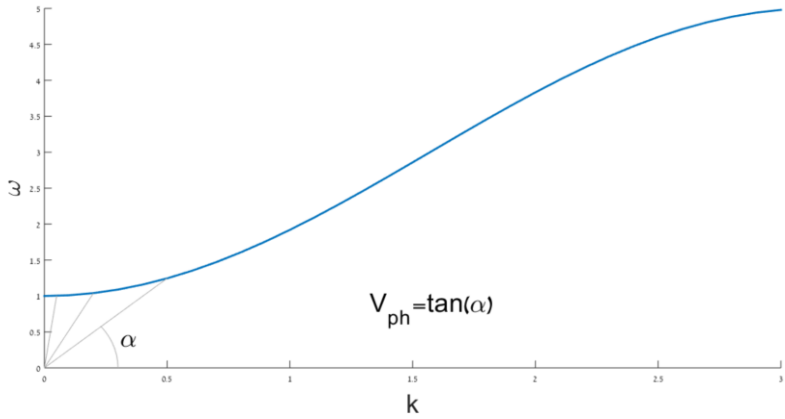

*Figure 8- A typical dispersion relation plot*

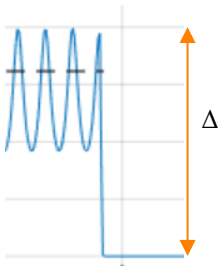

*Figure 9- Definition of $\Delta$*

$$\ddot{\varphi}+\beta\left[\varphi^3+(\varphi-\Delta)^3\right]\approx 0 \tag{7}$$

By substituting $z=\varphi/\Delta$, integrating over the entire motion range $0<z<1$ and by using $\tau\equiv\int_{t=0}^{t(\varphi=\Delta)}dt$, the following solution is obtained for the SDOF model:

$$\tau=\frac{\sqrt{2}}{\Delta\sqrt{\beta}}\int_0^1\frac{dz}{\sqrt{1-z^4-(z-1)^4}} \tag{8}$$

Evaluation of the integral and substitution $V=1/\tau$ further yields:

$$V=\frac{1}{\sqrt{2}\,\mathrm{K}\!\left(\frac{\sqrt{2}}{4}\right)}\Delta\sqrt{\beta} \tag{9}$$

Here $K$ is a complete elliptic integral of the first kind. Expression (9) can be represented decimally as $V=0.435\Delta\sqrt{\beta}$. Of course, the numeric coefficient in (9) should not be taken too seriously because of numerous approximations. The qualitative prediction to check is that the front velocity is related to the system parameters in the following way:

$$V\sim\Delta\sqrt{\beta} \tag{10}$$

However, it was found from numerical simulations that in order to obtain very good quantitative coincidence, it is enough to multiply (9) by constant coefficient $\gamma\approx 1.33$. Necessity of this correction stems from the simplification and assumptions taken; in the same time, with good accuracy, it is a constant, which does not depend on any of system parameters.

With this correction, one can also obtain an approximate equation for the time history of the particle inside the propagating front $z=z(\tilde{t})$:

$$\tilde{t}=\frac{\sqrt{2}}{\gamma}\int_0^z\frac{dz}{\sqrt{1-z^4-(z-1)^4}},\ 0<z<1 \tag{11}$$

Here the values of the actual time and $\varphi$ can be obtained from: $t=\left(\Delta\sqrt{\beta}\right)^{-1}\tilde{t},\ \varphi=\Delta z$.

Eq. (11) describes the movement of the considered single particle inside the front. The absence of system parameters in the rescaled representation implies that the basic structure of the kink does not depend on the system parameters and any specific solution is a rescaling of this basic shape. Comparison of this rescaled shape of the transition region to the one obtained numerically in the complete system is presented in Figure 10. The coincidence is reasonable, albeit not perfect.

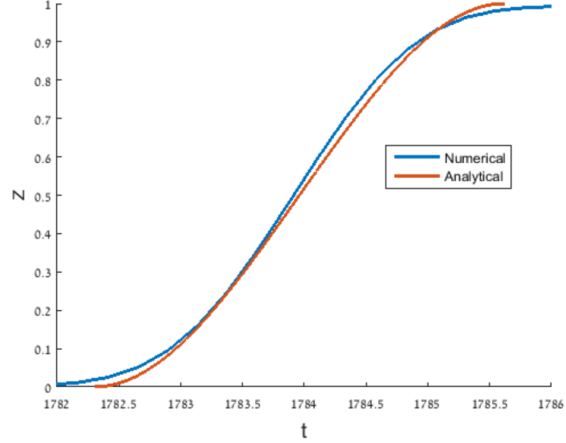

*Figure 10 – Comparison between analytic prediction and numeric result for the rescaled front shape.*

Kinetic relation (10) is even more important. It implies that the following relations should be satisfied with sufficient accuracy:

$$\begin{aligned} a) \quad & \ln V = 1 \cdot \ln(\Delta) + \ln\left(\gamma \left[\sqrt{2}\,\mathrm{K}\left(\frac{\sqrt{2}}{4}\right)\right]^{-1} \sqrt{\beta}\right) \\ b) \quad & \ln V = \frac{1}{2} \cdot \ln(\beta) + \ln\left(\gamma \left[\sqrt{2}\,\mathrm{K}\left(\frac{\sqrt{2}}{4}\right)\right]^{-1} \Delta\right) \end{aligned} \tag{12}$$

Direct comparison between the numerical and the analytical results for the front velocity are presented in Figure 11 and Figure 12 as functions of $\beta$ and $\Delta$ respectively.

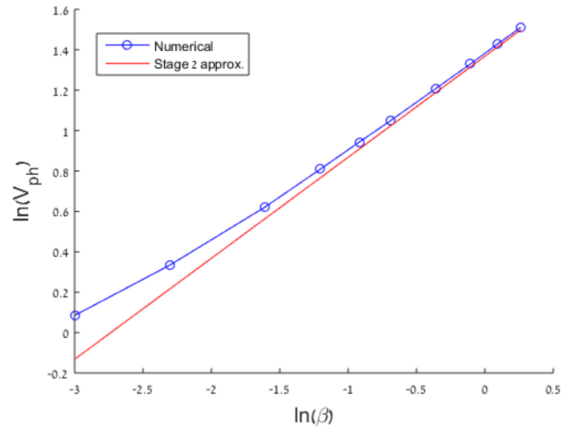

*Figure 11 – Validation of approximation of the nonlinear chain $\ln(V_{ph}) = f(\ln(\beta))$; Line-dotted blue – Numerical, Solid red – Analytical; parameters: $Q/B = 1, \omega_0 = 0.5$*

One of the main assumptions that led to derivation of the simplified model was the possibility to neglect the particularities of the on-site potential. The characteristics of the potential effected only the fixed value of the "amplitude of transition" $\Delta$. Each on-site potential yields different value of $\Delta$; however, the scaling law (13) that relates between the front velocity and $\Delta$ is valid for broad range of the on-site potentials with similar general characteristics ($B, Q, \varphi^*, b$). To check this claim, we examine three different potentials: bi-parabolic, 4$^{th}$ and 6$^{th}$ order polynomials with similar characteristics (see Figure 1). The comparison between these simulations and the analytic model is also presented in Figure 12. It is seen that all potentials provide same ratio between $\Delta$ and $V_{ph}$. This supports the assumption that the phase velocity depends only on the coordinate $\Delta$, rather than on the exact potential shape. This weak dependence leads to an important conclusion, that the model (10) is robust for modified shapes of the on-site potential. Also, the approximate model provides good estimation of the numerical results with accuracy that asymptotically improves at higher velocities.

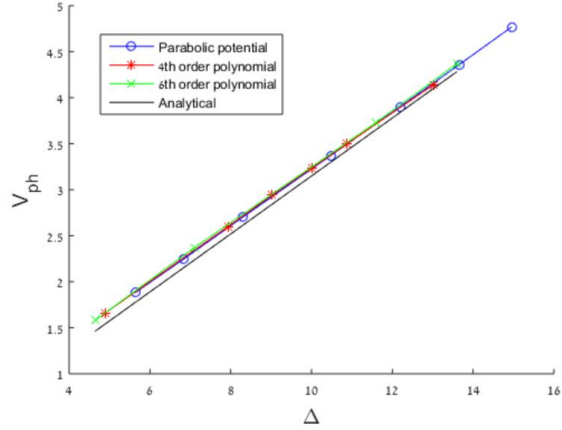

*Figure 12 – Front velocity as a function of potential parameter $\Delta$; 'o' blue – parabolic potential, '*' red – 4$^{th}$ order polynomial, 'x' green – 6$^{th}$ order polynomial; solid black – Analytical; Parameters: $\beta = 0.3, \omega_0 = 0.5$*

As it was demonstrated above, the front velocity linearly depends on $\Delta$. The latter parameter, however, is not known for a generic on-site potential, as it depends on the specific oscillation regime within the stable well. In the special case of bi-parabolic potential it is possible to evaluate this parameter explicitly. For this sake, we derive a nonlinear dispersion relation for the oscillatory tail. The piecewise parabolic potential has the following shape:

$$U(\varphi_n) = \begin{cases} \dfrac{\omega_0^2}{2}\varphi_n^2 & \varphi_n \leq b = \dfrac{\sqrt{2B}}{\omega_0} \\ \dfrac{\omega_1^2}{2}(\varphi_n - \varphi^*)^2 & \varphi_n > b \end{cases} \qquad (13)$$

Here, $\varphi^* = \dfrac{\sqrt{2(Q+B)}}{\omega_0} + \dfrac{\sqrt{2B}}{\omega_0}$. Let us introduce complex variable:

$$\Psi_n = \dot{\varphi}_n + i\omega\varphi_n \qquad (14)$$

The derivatives in terms of $\Psi$ are written as follows:

$$\varphi_n = -i\dfrac{\Psi_n - \Psi_n^*}{2\omega}, \quad \dot{\varphi}_n = \dfrac{\Psi_n + \Psi_n^*}{2}, \quad \ddot{\varphi}_n = \dot{\Psi}_n - \dfrac{i\omega}{2}(\Psi_n + \Psi_n^*) \qquad (15)$$

Substitution into the equations of motion for the stable branch and substitution of the modulated harmonic function: $\Psi_n = \phi_n e^{i\omega t}$, $\dot{\Psi}_n = \dot{\phi}_n e^{i\omega t} + i\omega e^{i\omega t}$ yields:

$$\dot{\phi}_n e^{i\omega t} + \frac{i\omega}{2}\phi_n e^{i\omega t} - \frac{i\omega}{2}\phi_n^* e^{-i\omega t} - \frac{i\omega_0^2}{2\omega}\left(\phi_n e^{i\omega t} - \phi_n^* e^{-i\omega t}\right) - \frac{i}{2\omega}\left[\left(2\phi_n - \phi_{n-1} - \phi_{n+1}\right)e^{i\omega t} - \left(2\phi_n^* - \phi_{n-1}^* - \phi_{n+1}^*\right)e^{-i\omega t}\right]$$
$$+ \frac{i}{8\omega^3}\beta\left[\left(\left(\phi_n - \phi_{n+1}\right)e^{i\omega t} - \left(\phi_n^* - \phi_{n+1}^*\right)e^{-i\omega t}\right)^3 + \left(\left(\phi_n - \phi_{n-1}\right)e^{i\omega t} - \left(\phi_n^* - \phi_{n-1}^*\right)e^{-i\omega t}\right)^3\right] = 0$$

(16)

Division by $e^{i\omega t}$ and subsequent averaging leads to the following slow-flow equations:

$$\frac{2\omega}{i}\dot{\phi}_n + \omega^2\phi_n - \omega_0^2\phi_n - \left(2\phi_n - \phi_{n-1} - \phi_{n+1}\right) - \frac{3}{4\omega^2}\beta\left[\begin{array}{c}\left(\phi_n - \phi_{n+1}\right)^2\left(\phi_n^* - \phi_{n+1}^*\right) + \\ + \left(\phi_n - \phi_{n-1}\right)^2\left(\phi_n^* - \phi_{n-1}^*\right)\end{array}\right] = 0 \quad (17)$$

By substituting $\phi_n = Ae^{ikn}$ one obtains the desired nonlinear dispersion relation:

$$\omega^2 - \omega_0^2 - 4\sin^2\frac{k}{2} + \frac{3\beta A^2}{\omega^2}\left[\sin^2 k - 4\sin^2\frac{k}{2}\right] = 0 \quad (18)$$

We see that the dispersion relation is indeed a small correction for the dispersion relation for linear chain (2). As it was mentioned above, the phase velocity of the oscillatory tail is large, consequently, the wavenumbers are near the left bandgap and the group velocity is small. Therefore, in the basic approximation the energy transport through the oscillatory tail can be neglected and the energy released due to the front propagation is almost not transferred towards or from the front. This situation strongly differs from one observed for the case of the linear coupling. [15]. The energy balance for arbitrary particle $n$ in the oscillatory tail can be simply expressed as:

$$E_n = \frac{\dot{\varphi}_n^2}{2} + \frac{\omega_0^2\left(\varphi_n - \varphi^*\right)^2}{2} \quad (19)$$

Due to the monochromatic structure of the tail, its form can be approximated as follows:

$$\varphi_n = \varphi^* + A\cos(kn - \omega t) \quad (20)$$

By substitution of (20) into (19) one obtains:

$$E_n = \frac{A^2}{2}\left[\omega_0^2 + \left(\omega^2 - \omega_0^2\right)\sin^2(kn - \omega t)\right] \quad (21)$$

We plug in the nonlinear dispersion relation (18) into (21). In the limit $k \to 0$ one obtains:

$$A \approx \frac{\sqrt{2Q}}{\omega_0} \quad (22)$$

In Figure 13 the amplitude is shown as a function of nonlinear coupling parameter $\beta$. The amplitude converges very quickly to the asymptotic value. At value of $\beta = 0.1$ the convergence is within 1% of the limit value.

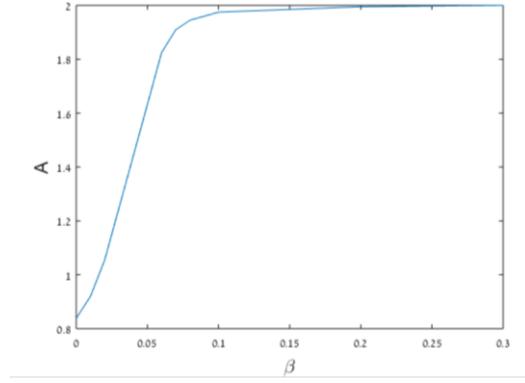

*Figure 13 - Amplitude of oscillation as a function of $\beta$ ; parameters: $Q=1, B=1, \omega_0 = 0.5$*

Thus, for the case of bi-parabolic on-site potential one obtains the explicit expression:

$$\Delta \approx \varphi^* + \frac{\sqrt{2Q}}{\omega_0} \qquad (23)$$

4. **Concluding remarks.**

First, the results presented above demonstrate the lack of robustness of the models with linear coupling. The addition of even small nonlinearity to the nearest-neighbor interaction drastically modifies the dynamic response and causes extremely high velocity and strong localization of the transition front. This effect can be attributed to combination of focusing properties of the gradient nonlinearity considered here, and the non-degenerate on-site potential. The latter provides constant supply of energy, which counterweights effects of pinning and radiation. As a result, extremely narrow kinks can propagate through the chain with far supersonic velocities.

Obviously, a continuous model of the system cannot explain such process. In the same time, namely the extreme localization and significant energy concentration in the front lead to considerable simplifications and reduction of the system to the effective single DOF model. Then, the dominance of the nonlinear term allows further simplifications and explicit prediction of the scaling relationships between the front velocity and other system properties: nonlinear coupling coefficient and maximal displacement of the particles inside the oscillatory front. These scaling relationships turn out to be universal for the on-site potentials with the same shape characteristics.

The authors are very grateful to Israel Science Foundation (grant 838/13) for financial support.